\documentclass[graybox]{svmult}

\usepackage{mathptmx}       
\usepackage{helvet}         
\usepackage{courier}        
\usepackage{type1cm}        
\usepackage{makeidx}         
\usepackage{graphicx}        
\usepackage{multicol}        
\usepackage[bottom]{footmisc}

\makeindex             

\begin{document}

\title*{The dependence of low redshift galaxy properties on environment}
\author{Simone M. Weinmann, Frank C. van den Bosch, Anna Pasquali}
\institute{S.M. Weinmann \at Leiden Observatory, Leiden University, 
P.O. Box 9513, 2300 RA Leiden, The Netherlands, \email{weinmann@strw.leidenuniv.nl}
\and F.C. van den Bosch \at Astronomy Department, Yale University, 
P.O. Box 208101, New Haven, CT 06520-8101, \email{frank.vandenbosch@yale.edu}
\and Anna Pasquali \at Astronomisches Rechen-Institut, M\"{o}nchhofstrasse 12-14, 
69120 Heidelberg, Germany, \email{pasquali@ari.uni-heidelberg.de}}
%
%
\maketitle

\vskip-1.2truein

\abstract{We review recent results on the dependence of various galaxy
  properties on environment at low redshift.  As environmental
  indicators, we use group mass, group-centric radius, and the
  distinction between centrals and satellites; examined galaxy
  properties include star formation rate, colour, AGN fraction, age,
  metallicity and concentration. In general, satellite galaxies
  diverge more markedly from their central counterparts if they reside
  in more massive haloes.  We show that these results are consistent
  with starvation being the main environmental effect, if one takes
  into account that satellites that reside in more massive haloes and
  at smaller halo-centric radii on average have been accreted a longer
  time ago.  Nevertheless, environmental effects are not fully
  understood yet. In particular, it is puzzling that the impact of
  environment on a galaxy seems independent of its stellar mass. This
  may indicate that the stripping of the extended gas reservoir of
  satellite galaxies predominantly occurs via tidal forces rather than
  ram-pressure.}

\section{Introduction}

It has long been known that galaxies living in dense regions tend to
be redder, less active in their star formation and of earlier type
\cite{mb97, ad80, ao74}. More recently, large galaxy surveys like the
SDSS and reliable stellar mass estimates have indicated that the main
parameter governing galaxy properties is however not environment, but
stellar mass \cite{gk03}. It is therefore crucial that environmental
effects are studied at fixed stellar mass.

Traditionally, environment has been described in terms of galaxy
density (for example out to the $n$-th nearest neighbour), or in terms
of the field versus cluster distinction.  Group and cluster finding
algorithms applied to the SDSS and other surveys have now made it
possible to quantify environment in a way that is physically better
motivated, expressing it in terms related to the expected underlying
dark matter structure \cite{sb09, sh09, va07,xy07}.  To first order,
one can quantify environment simply by discriminating between central
and satellite galaxies \cite{fv08b}, which acknowledges the fact that
those two kinds of galaxies have different average dark matter
accretion histories. Finer distinctions between different
subpopulation of satellites can then be made according to their host
halo mass and group- or cluster-centric distance.

Taking out stellar mass dependencies, and describing environment with
a physically better motivated language makes it easier to interpret
the observed environmental dependencies. Many different processes have
been suggested to drive these environmental dependencies, including
ram-pressure stripping of the cold gas \cite{jg72}, removal of the
extended (hot) gas reservoir of galaxies by tidal or ram-pressure
stripping ('starvation', \cite{rl80, mb97}), harassment by high-speed
tidal encounters \cite{bm96}, or a faster mass growth at early times
for galaxies that end up in more massive haloes at late times
\cite{en10}.  Much of the recent work on this topic converges towards
``starvation'' being the main driver of environmental effects
\cite{sw06a, va10,vdl10}. This is an outcome foreseen by early
semi-analytical models which included no other environmental effect
except the instantaneous removal of the satellite's hot gas
reservoir at accretion \cite{gk93}. 
However, it has also been shown that this simple recipe 
likely makes starvation
in semi-analytical models overefficient \cite{sw06b, im08}, and there are still 
open questions on how
and on which timescales this effect exactly operates.

In this review, we summarize recent results on how various galaxy
properties depend on environment at fixed stellar mass. We then
outline potential implications of these results on our understanding
of galaxy evolution as a function of environments.

\section{SDSS group and cluster catalogues}

Most of the results described in what follows are based on the Yang et
al. \cite{xy07} group catalogue, which makes use of the SDSS-DR4
\cite{ja06} and the New York Value-Added Galaxy Catalogue
\cite{mb05}. This group catalogue is constructed using the halo-based
group finder of \cite{xy05}, which uses an iterative scheme and priors
on the redshift-space structure of dark matter haloes to partition
galaxies over groups. Halo masses are estimated from a ranking in
total characteristic luminosity or total characteristic stellar mass,
with both methods giving very similar results \cite{xy07}. Details on
the galaxy group sample used in most of the studies mentioned below
can be found in \cite{fv08a}.  Some results come from the cluster
catalogue by von der Linden et al. \cite{va07}, which is based on the
SDSS DR4 catalogue and the C4 cluster catalogue \cite{cm05}.  In what
follows, a ``central'' galaxy is defined as the most massive galaxy in
its group. All other galaxies which reside in the same group are
labelled ``satellites''.

\section{Results}
\subsection{The dependence of galaxy properties on the satellite-central dichotomy}
\begin{figure}[b]
\sidecaption[t]
\includegraphics[scale=.70]{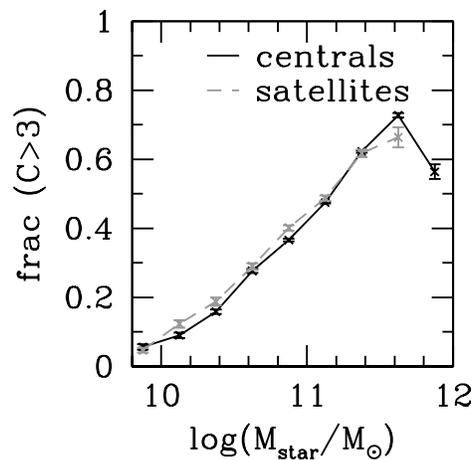}
\caption{The fraction of galaxies with a SDSS concentration greater
  than 3 for central galaxies (solid line) and satellites (dashed
  line).  Concentration is defined as the ratio between the radius
  containing 90 percent of the r-band light, divided by the radius
  containing 50 percent of the r-band light. More details are given in
  \cite{sw09}.}
\label{fig:w09}       
\end{figure}

To first order, we can quantify environmental dependence simply by
comparing satellite and central galaxies of the same stellar mass.
Several key differences between those two types of galaxies have been
found.  The most basic difference is that satellite galaxies are
redder \cite{fv08a} and have lower specific star formation rates
\cite{tk09} than central galaxies of the same stellar mass. In
addition, satellite galaxies are less likely to reveal optical or
radio AGN activity, and their AGN activity is weaker than in their
central counterparts \cite{ap09}. In terms of morphology, it is found
that satellite galaxies have surface brightness profiles that are, on
average, slightly more concentrated than those of central galaxies
\cite{fv08a, sw09, yg09}. As shown by \cite{sw09}, this most likely
does not reflect a true difference in the mass distribution, but
rather can be explained by fading of the stellar disk due to star
formation quenching in the satellites. This is consistent with the
finding that satellites and centrals reveal no structural differences
if they are matched in both stellar mass and colour \cite{yg09,
  fv08a}. As can be seen in Fig.~\ref{fig:w09}, the fraction of
galaxies with a concentration greater than 3 is the same for
satellites and centrals at fixed stellar mass. This suggests that the
most highly concentrated galaxies (which typically have elliptical
morphologies) are not produced by environmental effects.

\begin{figure}[b]
%
\includegraphics[scale=.70]{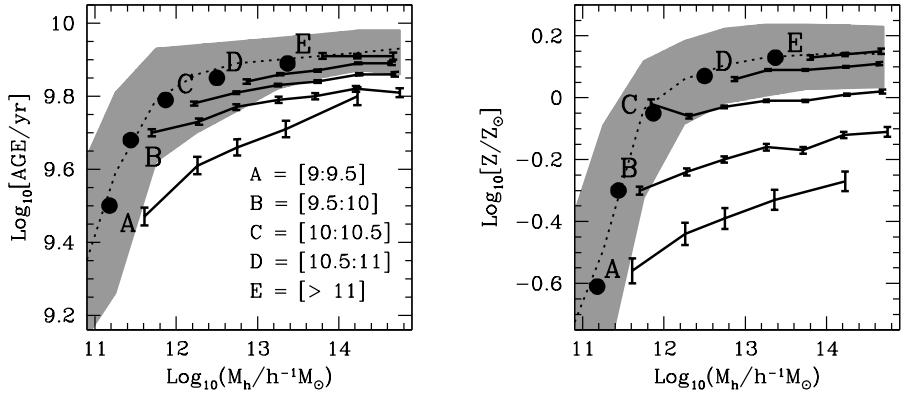}
\caption{The mean stellar mass-weighted age (left hand panel) and
  metallicity (right hand panel) for central galaxies (large filled
  circles) and for satellite galaxies (lines with errorbars) for
  different stellar mass bins A-E, as indicated. Results for
  satellites are shown as a function of halo mass. It can be seen that
  low mass centrals have lower metallicities than satellites, and that
  the metallicity for low mass satellites increases with increasing
  host halo mass. More details are given in \cite{ap10}.}
\label{fig:pasquali1}       
\end{figure}

\subsection{The dependence of satellite properties on  host halo mass}

Group catalogues make it possible to study the properties of satellite
galaxies as a function of the mass of their host halo.  While the
average colours of satellites depend only very weakly on host halo
mass \cite{fv08b}, there seems to be a significant increase of the
fraction of red or passive satellites with increasing halo mass
\cite{tk09, va10}.  Interestingly, satellites in more massive haloes
are older and more metal rich than their counterparts in lower mass
haloes \cite{ap10}. This is shown in Fig.~\ref{fig:pasquali1}, which
plots the stellar mass-weighted ages and metallicities of centrals and
satellites as a function of host halo mass. The halo mass dependence
(and hence the difference between centrals and satellites) is most
pronounced at relatively low stellar masses, and disappears at the
high mass end \cite{ap10}.

While the older ages of satellites are likely directly related to
their larger passive fraction, their higher metallicity is less
straightforward to explain. It could indicate that the satellites are
the descendants of centrals with higher stellar mass, and thus a
higher metallicity, which underwent tidal stripping of their stellar
material \cite{ap10}. However, the amount of stellar mass stripping
required is fairly substantial, which is difficult to reconcile with
the fact that satellites seem to have the same concentrations like
centrals of the same stellar mass (see for example
\cite{cm05}). Another explanation could be that satellites form a
significant amount of stars after infall. It is plausible that these
would have a higher metallicity than stars formed in a comparable
central galaxy due to the lack of new infall of low metallicity
(`primordial') gas. However, a significant boost in the stellar
mass-weighted metallicity requires that satellites form a relatively
large fraction of their stars after accretion, which may proof
difficult to reconcile with their high passive fractions. Clearly the
origin of the relatively high metallicities in low mass satellites
needs to be investigated in more detail, and might give interesting
new insights into the processing and recycling of gas in both central
and satellite galaxies.

Finally, although there are indications from HOD modelling that there
is a relation between the fraction of satellites with AGN activity and
host halo mass \cite{tm10}, no such relation has been found using our
group catalogue \cite{ap09}.

\subsection{The dependence of satellite properties on cluster-centric radius}

\begin{figure}[b]
\sidecaption[t]
\includegraphics[scale=.70]{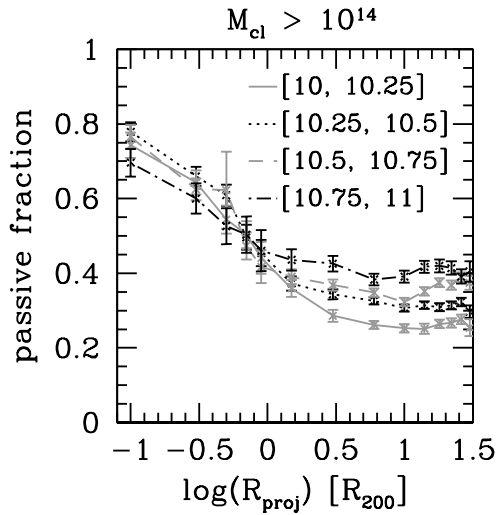}
\caption{The fraction of passive satellites in clusters with halo
  masses $M >10^{14} h^{-1}M_{\odot}$, obtained from \cite{va07}, as a
  function of projected cluster-centric radius. Results are shown for
  four different stellar mass bins, as indicated [values in brackets
  refer to $\log(M_{\rm star}/h^{-2}M_{\odot})$]. Here `passive' is
  defined as having a SSFR $< 10^{-11} {\rm yr}^{-1}$ Note the clear
  decrease with increasing cluster-centric radius. For more details
  see \cite{sw10}.}
\label{fig:weinmann2}       
\end{figure}

Dependencies of galaxy properties on group- or cluster-centric
distance are most clearly visible in massive clusters where statistics
are best and the center of the cluster is easier to define than in
poor groups.  Fig.~\ref{fig:weinmann2} shows the fraction of passive
satellites in haloes with $M > 10^{14} h^{-1} M_{\odot}$ as a function
of halo-centric radius in the sample used by \cite{sw10}.  Clearly, in
these massive haloes the passive fraction of satellites at a given
stellar mass increases towards the center \cite{vdl10, sw10, sb09}.
Interestingly, the fraction of galaxies showing signs of {\it fast
  recent} truncation of star formation is found to be virtually
independent of cluster-centric radius \cite{vdl10}, which seems to
suggest that {\it fast} truncation may be unrelated to environmental
effects.  Finally, the fraction of galaxies hosting a powerful optical
AGN has been found to decrease towards the cluster center at fixed
stellar mass \cite{vdl10}.

\subsection{The puzzling independence of environmental effects on stellar mass}

\begin{figure}[b]
\sidecaption[t]
\includegraphics[scale=.70]{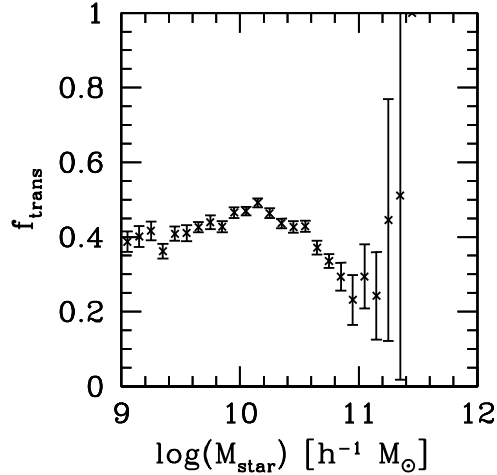}
\caption{The transition fraction $f_{\rm trans}$, which expresses the
  blue-to-red transition fraction of satellites that were still blue
  at the time of accretion (see Eq.~[\ref{ftrans}]), as a function of
  stellar mass. A galaxy is defined as red if $^{0.1}(g-i) > 0.76 +
  0.15 \cdot$ [log$(M_{\rm star}/h^{-1}M_{\odot})-10.0$]. See
  \cite{fv08a} for details.}
\label{fig:vdb1}       
\end{figure}

An interesting question that has not received much attention yet is
how the impact of environment depends on the stellar mass of the
galaxy. Naively, one would expect that stripping of the hot or cold
gas in a galaxy by any kind of effect is easier for a shallower
potential well, and therefore that low mass galaxies are more
vulnerable to environmental effects.  To investigate the dependence of
the strength of environmental effects on stellar mass, we can
introduce a quantity $f_{\rm trans}$, which gives us an estimate on
the fraction of galaxies which were blue at the time of infall and
have by now become red due to environmental effects \cite{fv08a}:
\begin{equation}\label{ftrans}
 f_{\rm trans} = \frac{f_{\rm sat, red}  - f_{\rm cen, red}}{f_{\rm cen, blue}}
\end{equation}
with $f_{\rm sat, red}$ and $f_{\rm cen, red}$ the red fraction of
satellites and centrals respectively, and $f_{\rm cen, blue}$ the blue
fraction of centrals. Note that we assume here that the blue fraction
of centrals today corresponds to the blue fraction of the satellites
at the time of infall, which should be roughly correct, since most
satellites fall in relatively late \cite{fv08a}. As shown in
Fig.~\ref{fig:vdb1}, $f_{\rm trans}$ is remarkably constant at around
40 \% from $10^9 M_{\odot}$ up to $10^{11} M_{\odot}$.  A similar
result was obtained by \cite{yp10}.  These findings imply that the
probability for a galaxy to become red due to its environment is
nearly independent of its stellar mass, which is challenging to
understand from a theoretical perspective.

\section{Discussion}

The fact that satellites and centrals are different is not surprising
given that the former are believed to reside in dark matter subhaloes
orbiting within a larger dark matter halo (the `host' halo), while
centrals are expected to reside at rest at the center of a host halo.
Whereas host haloes continue to grow in mass via accretion, subhaloes
lose mass (to their host) due to tidal stripping. This implies that
subhaloes, and their associated satellites, also lack the accretion of
intergalactic gas associated with halo growth.
However, this effect by itself does not necessarily mean that centrals
and satellites are different. After all, if satellite galaxies can
hang on to their extended haloes of hot gas which they are predicted
to have at infall, they may remain `active' for a relatively long
period of time \cite{sw10}. An additional environmental effect
therefore seems required to explain the observed dependencies. Based
on all results discussed above, this effect has to have the following
properties:
\begin{itemize}
\item It is likely to result in a depletion of gas, which directly
  causes higher passive fractions, lower AGN fractions, and higher
  average stellar-mass weighted ages in satellites compared to
  centrals of the same stellar mass.
\item It occurs for satellite galaxies in haloes spanning a large
  range in mass (not only in clusters), but becomes weaker towards
  less massive haloes, and towards the outskirts of clusters \cite{tk09,
    vdl10, sw10, sb09}.
\item It quenches star formation in galaxies on a rather long
  timescale of the order of 2-3 Gyr \cite{xk08, sw09, vdl10, va10}.
\item There are no indications that it results in an accompanying
  structural transformation \cite{sw09, va08, yg09}.
\item It is similarly strong for galaxies with different stellar
  masses \cite{fv08a, yp10}
\end{itemize}
Of all the environmental effects that have been suggested, starvation
matches this description best. This effect should take place in all
kinds of groups, not only in massive clusters. Its effect on the star
formation rate is slow, and it does not lead to any morphological
changes in the galaxy.  However, how this effect operates in detail is
still under debate. Also, it is not a priori clear why it should be
similarly strong for galaxies with different stellar masses.

Semi-analytical models that include instantaneous and complete removal
of the extended gas reservoir of satellites upon accretion result in a
passive or red fraction of satellites that is much too high
\cite{sw06b}. It has therefore become clear that starvation has to be
a more gradual process, which is taken into account in some of the
newest semi-analytical models \cite{af08, sw10, qg10}.  However, there
are indications that modelling ram-pressure stripping of the hot halo
according to standard prescriptions leads to a too high fraction of
low mass, passive galaxies \cite{sw10} and also produces too many
satellite galaxies with intermediate colours \cite{mb97, sw10}.  This
could indicate that starvation (i.e. the depletion of the hot gas
reservoir) mainly occurs by tidal stripping, and not by ram-pressure
stripping \cite{sw10}. Another argument for this is that tidal
stripping seems to be less strongly stellar mass dependent than
ram-pressure stripping \cite{sw10}, which likely helps in reproducing
the stellar-mass independence of environmental effects discussed
above.  It also does not seem entirely unrealistic that ram-pressure
stripping could be overestimated in standard semi-analytical models,
as these tend to overestimate the hot gas content of groups
\cite{rb08, sw10}.

\begin{figure}[b]
%

\includegraphics[scale=.60]{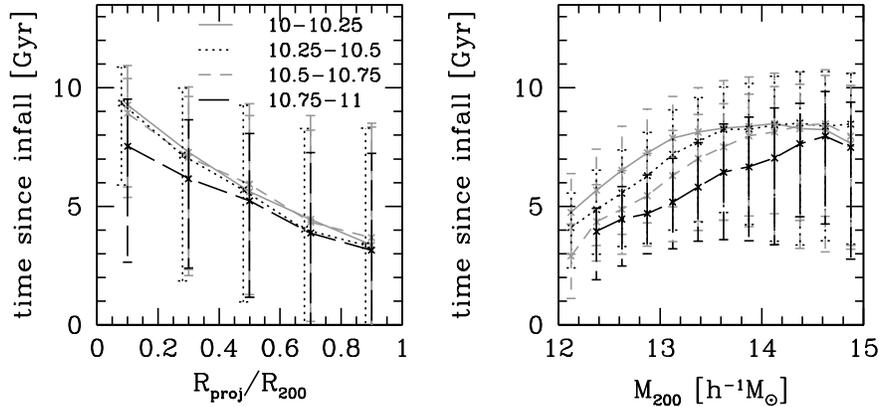}
\caption{The median time since galaxies switched from the central to
  the satellite status for the last time in a semi-analytical model
  \cite{gd07}, for four different stellar mass bins, as indicated. In
  the left hand panel, we use a sample of clusters selected in a
  similar way as in observations, as described in detail in
  \cite{sw10}, and plot the median time since infall as a function of
  projected cluster-centric distance. Cluster masses are estimated
  according to velocity dispersion, as explained in \cite{sw10}. We
  only use clusters with ``observed'' masses of $10^{14} - 10^{15}
  h^{-1}M_{\odot}$.  In the right hand panel, we plot the median time
  since infall as a function of group mass, for groups as selected
  directly in the Millennium simulation.  Errorbars in both panels
  denote the range where 68 \% of the measured values lie.  }
\label{fig:weinmann1}       
\end{figure}

The observation that galaxies in cluster centers and in more massive
groups have a higher passive fraction than their counterparts in the
outskirts and in less massive groups can have two different
implications. Either environmental effects are simply strongest in
group centers and in massive clusters, or galaxies residing at smaller
halo-centric distances and/or in more massive haloes have on average
been satellites for a longer period so that environmental effects have
had more time to operate.  Indeed, for the semi-analytical model of
\cite{gd07}, we find that satellites in more massive systems, and
closer to the projected cluster-center, have been satellites for
longer period, which is shown in Fig. \ref{fig:weinmann1}. Neither
dependency is surprising. Galaxies that end up in more massive haloes
are `born' in denser environments, and therefore tend to become
satellites at earlier times. Also, it takes time for newly infalling
galaxies to sink to the cluster center by dynamical friction, which
explains the radial dependence (see also \cite{lg04, ga10}). Hence,
based on Fig.~\ref{fig:weinmann1}, we conclude that the larger
fractions of passive satellites in cluster centers and in massive
clusters most likely reflect trends in the time of infall, and do
not require that environmental effects are stronger in more
massive haloes and/or at smaller halo-centric radii.

\section{Outlook}

Models of environmental effects start to be able to reproduce the
relations between galaxy properties and environment for low redshift
galaxies with masses above $10^{9} M_{\odot}$ \cite{sw10}, indicating
that our understanding of environmental effects is improving.  Of
course, these models should be further tested and refined by using
more detailed observational data at low redshift. For example, the gas
content of galaxies in different environments can hold important
additional clues on the interplay between accretion, SN feedback and
environmental effects \cite{gk10}.

Another important test for these models is whether or not they can
capture the redshift evolution of environmental dependencies.
However, at high redshift there are still several important
discrepancies between current semi-analytic model predictions and
galaxy properties in general \cite{if09, ff09, qg10} which may need to
be addressed first.

Finally, it is important to probe environmental effects at masses
lower than discussed here. Although the efficiency of environmental
effects seems to be nearly independent of stellar mass in the mass
range discussed here, this might well change at even lower stellar
masses. To explain the observed population of dwarf elliptical
galaxies and their different subclasses as found for example in the
Virgo cluster \cite{tl07}, processes additional to starvation, like
harassment, or a different formation channel at early times, might be
required.

\begin{acknowledgement}
  We thank all our collaborators on this topic, in particular Houjun
  Mo, Xiaohu Yang, Guinevere Kauffmann, Anja von der Linden, Gabriella
  De Lucia, Anna Gallazzi, Fabio Fontanot, Yicheng Guo, Daniel
  McIntosh, and Xi Kang.
\end{acknowledgement}
%


\end{document}